\documentstyle[aps,prc,epsfig,preprint]{revtex}

\tighten
\begin{document}
\draft
\title{Shell  Corrections  for Finite-Depth Deformed Potentials:\\
Green's Function Oscillator Expansion Method}
\author  {T. Vertse,$^{1,2}$
           A.T. Kruppa,$^{1,2}$
           W. Nazarewicz,$^{3-5}$
           }
\address {
          $^1$Institute of Nuclear Research of the Hungarian
	  Academy of Sciences \\ 
          P.O. Box 51, H-4001, Debrecen, Hungary \\[1mm]
$^2$Joint Institute for Heavy Ion Research,
              Oak Ridge National Laboratory,
              P.O. Box 2008, Oak Ridge,   Tennessee 37831\\[1mm]  
         $^3$Department of Physics and Astronomy, University
          of Tennessee,
Knoxville,  Tennessee 37996\\[1mm]
$^4$Physics Division, Oak Ridge National Laboratory,
   P.O. Box 2008\\
 Oak Ridge,   Tennessee 37831\\[1mm]
          $^5$Institute of Theoretical Physics, Warsaw University,
	  ul. Ho\.za 69, PL-00681, Warsaw, Poland
}

\maketitle
\begin{abstract}
Shell corrections of the finite deformed Woods-Saxon  potential are
calculated using the Green's function method and the generalized Strutinsky 
smoothing procedure. They are compared  with the results of the
standard prescription   which are affected by the spurious 
contribution from the unphysical particle gas.
In the new method,  the shell correction approaches the  exact limit
provided that  the dimension of the single-particle
 (harmonic oscillator) basis is sufficiently large. For spherical potentials,
  the present method is
faster than the exact one in which the contribution from the
particle continuum states is explicitly
calculated. For deformed potentials,
the Green's function method offers a practical and reliable way of calculating shell corrections 
for weakly bound nuclei.
\end{abstract}


\narrowtext

\section{Introduction}

The positive-energy  spectrum of the average single-particle
potential plays a role  in  the description of weakly bound nuclei 
for which the Fermi level
approaches zero (see Ref.~\cite{[Dob97a]}).
 For these nuclei, important  for both nuclear structure and
nuclear astrophysics studies, special  care should be taken when  dealing
with the particle continuum  which  seriously impacts many
nuclear properties, including bulk nuclear properties  (e.g., masses, radii, shapes) as well
as nuclear dynamics (i.e., excitation modes).

In two earlier papers \cite{[Naz94],[Ver98]}, a
macroscopic-microscopic  method was
applied
to nuclei far from the beta stability line.
It has been demonstrated that the
 positive-energy single-particle  spectrum
does impact the results significantly, 
 and  the systematic error
in binding energies,
due to the neglect or the improper treatment  of the particle continuum,
can be as large as several MeV at the neutron drip line.
In the first paper \cite{[Naz94]},  both
spherical and deformed nuclei were considered, and the continuum
was approximated by a limited number of quasistationary states 
which resulted
from a diagonalization of the Woods-Saxon average 
potential in a harmonic oscillator basis.
In the vicinity of the neutron drip line,  the result of the Strutinsky
smoothing (standard averaging method) becomes unreliable  and it deviates
from the result of the semiclassical Wigner-Kirkwood expansion.
The semiclassical method, which does not use the positive-energy spectrum explicitly, 
gives a more reliable estimate of the shell correction than the standard method
(see Refs.~\cite{[Ros72],[Naz94]} and references quoted therein).

In the following  study \cite{[Ver98]}, carried out
 for spherically symmetric nuclei,  a more detailed comparison was carried out
between the Strutinsky  smoothing method  and the  Wigner-Kirkwood expansion.
In the Strutinsky method, the continuum effect
was taken into account  exactly,  i.e., by calculating the
continuum part of the level density from the derivative of the scattering
phase shift with respect to single-particle energy. The smooth part of the continuum
level density has been calculated by means of the 
contour integration along a path in the complex energy plane
\cite{[San97a]}. 
Although the continuum level density was treated
properly,  it has been concluded  \cite{[Ver98]} that in most nuclei the 
plateau condition of the Strutinsky method \cite{[Bra72]}
could not be met. Therefore,  this condition was replaced
with the requirement of the linear energy dependence of the mean level density.
This modification (which widens the range of the applicability of the Strutinsky
procedure  considerably) is referred to as  the {\em generalized Strutinsky method}. 
The new procedure 
has  proved to be very useful; in most cases
it gives results  reasonably close to the estimate of the semiclassical method.
The exceptions are the neutron drip-line nuclei in which
the neutron Fermi level approaches zero and
the semiclassical procedure diverges \cite{[Ver98]}. 

In the present work, the effect of the particle continuum on  shell
correction is further studied for both spherical and deformed nuclei
by using the recently introduced
Green's function approach \cite{[Kru98],[Shl97]}. The advantage of this method 
is that, by  diagonalizing
the finite single-particle potential in a square-integrable basis,
one  can get rid of the
spurious contribution of the particle gas to the level density.
We shall refer to this novel method as the {\em new
method} 
in order to distinguish it from the commonly used  standard smoothing
procedure   (or: {\em old method})
 in which the spurious 
contribution from  the particle gas is  not subtracted (but diminished  by using a reduced
number of basis states). 

The paper is organized as follows. Section~\ref{shellcorr}
contains a  brief review of the shell-correction method  and 
describes several  versions of the smoothing procedure used.
The numerical results of shell-correction  calculations are presented in Sec.~\ref{results}
for both neutron-rich and
proton-rich nuclei.
Finally, conclusions are drawn  in Sec.~\ref{conclusions}.

\section{Strutinsky Smoothing Procedure}\label{shellcorr}

\subsection{Basic Definitions}\label{basicdefs}

In the macroscopic-microscopic approach \cite{[Bra72],[Str67],[Str68],[Bol72],[Mol81]},
 the shell correction, 
\begin{equation}\label{e1}
 \delta E_{\rm shell} = E_{\rm s.p.} - \tilde{E}_{\rm s.p.},
\end{equation}
 is defined as the difference between  the
total single-particle
 energy  $E_{\rm s.p.}$,
\begin{equation}\label{shell}
E_{\rm s.p.} =
\sum_{i-{\rm occ}}\epsilon_i,
\end{equation}
and  the  smooth single-particle  energy  $\tilde{E}_{\rm s.p.}$.
The shell correction represents the fluctuating part of the binding
energy resulting from the  quantal single-particle shell structure.

For the sake of simplicity,  we shall  assume that the single-nucleon
energy spectrum is  that of a
one-body  Hamiltonian,
\begin{equation}\label{Hamiltonian}
 \hat{H}=\hat T+\hat V,
\end{equation} 
with  an average
single-particle potential $\hat V$.
In practice,  the potential  contains a 
deformed Woods-Saxon potential, the
spin-orbit term and the  Coulomb potential.
Since the central potential
 is finite, the spectrum of $\hat H$ is composed of bound states
with discrete negative eigenvalues
($\epsilon_i<0$) and the  continuum of scattering states with positive energies
 ($\epsilon>0$). Consequently, 
the single-particle level density is
\begin{equation}\label{e3}
g(\epsilon) = g_{\rm d}(\epsilon) + g_{\rm c}(\epsilon),
\end{equation}
where 
\begin{equation}\label{discreteleveldens}
g_{\rm d}(\epsilon)=\sum_i 2~ \delta(\epsilon-\epsilon_i)
\end{equation}
is the level density of the discrete (bound) states and $ g_{\rm c}(\epsilon)$  is 
the  continuum level density (it will be specified later).
The factor $2$ in Eq.~(\ref{discreteleveldens}) appears due to the two-fold
Kramers degeneracy of the deformed single-particle energy levels.

In the shell-correction method
\cite{[Str67],[Str68]},   $\tilde{E}_{\rm s.p.}$
is calculated by employing  the
smoothed  level density $\tilde{g}(\epsilon)$ obtained from $g(\epsilon)$
by folding it with a smoothing function
$f(x)$:
\begin{equation}\label{ggsmooth}
\tilde{g}(\epsilon) = {1\over\gamma}
\int_{-\infty}^{+\infty}d\epsilon'~ g(\epsilon')
f\left(\frac{\epsilon'-\epsilon}{\gamma}\right).
\end{equation}
In practical applications,  for the folding function $f(x)$ one is usually 
taking  a product of a
Gaussian weighting  function, $\frac{1}{\sqrt{\pi}}\exp(-x^2)$,  and a
corresponding curvature correction
polynomial of the  order $p$ \cite{[Str68]} which is an associated Laguerre
polynomial $L_{p/2}^{1/2}(x)$ ($p$-even).  
The smoothed level density (\ref{ggsmooth}) defines both the
 smooth single-particle  energy
\begin{equation}\label{Esmoth}
\tilde{E}_{\rm s.p.}=\int_{-\infty}^{\tilde\lambda}
\epsilon\, \tilde{g}(\epsilon) d\epsilon,
\end{equation}
and  the smoothed Fermi level $\tilde{\lambda}$. The latter is
obtained  from
the particle number equation:
\begin{equation}\label{ltilde}
N = \int_{-\infty}^{\tilde{\lambda}} \tilde{g}(\epsilon)d\epsilon.
\end{equation}

The smooth single-particle energy and the resulting shell correction
have to be defined unambigously.
Therefore, they must  neither  depend  on the smoothing range $\gamma$
nor on the order $p$ of the  curvature correction.
This requirement,  referred to as the   {\em plateau condition}, 
can be written as
\begin{equation}\label{plateau}
\frac{d\tilde{E}_{\rm s.p.}}{d\gamma}=0\ \ ,
\frac{d\tilde{E}_{\rm s.p.}}{dp}=0.
\end{equation}
Of course, since one  wants to eliminate  the oscillations due to the shell structure,
the smoothing range $\gamma$   should be
greater  than  the average energy distance between
neighboring major shells,  $\hbar\omega_0$$\approx$41/$A^{1/3}$\, MeV
\cite{[Boh69]}. 

For  infinite potentials, such as an infinite square well, 
harmonic oscillator,  and  a deformed Nilsson potential,
one can always find a range 
of the smoothing parameters $\gamma$ and $p$ in which the
 smooth single-particle energy is independent of the
values  $\gamma$ and $p$ \cite{[Bra73],[Ros72]}.
For  finite-depth  potentials, however, additional complications
 arise due to  (i) the presence of
 positive-energy continuum and (ii) the difficulties with meeting the plateau condition.
 We shall discuss these points in the following.

\subsection{Effect of the Unbound Spectrum}

The need for calculating the continuum level density, $g_{\rm c}(\epsilon)$,
 appears whenever one  deals with
 finite-depth potentials.
For spherically symmetric potentials,  the continuum level density  
\cite{[Ros72],[Shl92],[Uhl36],[Bet37],[Li70]} is defined by means of the
scattering phase shifts $\delta_{lj}(\epsilon)$:
\begin{equation}\label{e4}
g_{\rm c}(\epsilon)={1\over \pi}
\sum_{l,j}(2j+1)\frac{d\delta_{lj}(\epsilon)}{d\epsilon}.
\end{equation}
For realistic nuclear potentials,  phase shifts have to be calculated by numerically 
solving the radial Schroedinger equations and by matching the wave function
to the asymptotic solution at a distance where the nuclear potential
can be neglected. This procedure has to be carried out  for every  partial wave below
a certain angular momentum cut-off  on a fine mesh in the positive-energy
region.   In order to prevent sudden jumps in $g_{\rm c}(\epsilon)$ around narrow resonances, 
a new calculational method  employing the Cauchy theorem was introduced 
in Ref.~\cite{[San97a]}. Here, 
the complex energies $w_i$ of the Gamow resonances 
(poles of the S-matrix) are localized first, then a contour
of the complex energy plane is chosen. The contour,  denoted by $L$, should go far 
away from 
the poles.  The mean level density $\tilde g(\epsilon)$ is then  
calculated as a sum  over  bound and those  resonant states which  lie  between $L$ and the real
energy axis  and
an integral term along a contour:
\begin{equation}\label{gcsmooth}
\tilde{g}(\epsilon) =
\sum_{i}f\left(\frac{\epsilon-w_i}{\gamma}\right)
+\int_{L}dw~g_{\rm c}(w)
f\left(\frac{\epsilon-w}{\gamma}\right).
\end{equation}
Apart from the
numerical errors,
this procedure gives the continuum level density exactly. Therefore,
we call this approach as {\em numerically exact} or, simply,   {\em
exact}.

For  deformed single-particle potentials, the continuum level density has a more complicated
form and can be expressed by the on-shell S-matrix $S(\epsilon,\hat k,\hat k^{ \prime})$ as
\cite{[Kru98]}
\begin{equation}\label{d4}
g_{\rm c}(\epsilon)={1\over {2i\pi}}{\rm Tr}[S(\epsilon,\hat k,\hat
k^{\prime})^*{d\over {d\epsilon}}S(\epsilon,\hat k,\hat k^{\prime})].
\end{equation}
If one  wants to use  expression   (\ref{d4}) for calculating the continuum level density,
one has to determine the S-matrix by solving the coupled
system of differential equations for each value of  $\epsilon$
 \cite{[Kru98]}. In practice this is  a  difficult task.

In the old method, the single-particle Hamiltonian is  diagonalized
in a square-integrable basis formed from the eigenstates of an 
infinite potential. This
 potential can be  either a finite-range potential 
contained in an impenetrable box (i.e., an infinite wall at a certain
distance)  or the  harmonic oscillator potential. Since the number
of basis states is always assumed to be finite,  the diagonalization results in a discrete set of
eigenstates. The eigenstates with negative energy approximate the bound states
of the original Hamiltonian, while the positive-energy quasi-bound states
mock-up  the effect of the particle continuum in a very crude way.
It has  early been  realized  that in the application of this method one should
not use too large a
basis; otherwise,  the level density around the zero energy would increase dramatically.
(In fact it diverges as the basis size  goes to infinity.) In order to
avoid this  catastrophe,  the use of a harmonic oscillator basis with
12-14 harmonic oscillator shells was recommended \cite{[Bol72]}.
One of the objectives  of this paper is to perform the critical evaluation 
of the standard smoothing method by calculating the 
continuum level density in a more reliable  way using the Green's function
approach.

\subsection{Green's Function Method}

Although the Green's function approach to the single-particle level
density was developed long ago  (see, e.g.,
Refs.~\cite{[Shl92],[Bau83],[Bal70],[Bal71]}), it is  somehow surprising that
so far it  has not been widely applied. 
Below, we briefly summarize the main features of this method. More details
can be found in Refs.~\cite{[Kru98],[Shl97]}.

A Hamilton operator with an infinite potential, $\hat H_{\infty}$, has only 
discrete energy eigenvalues  and its  eigenfunctions are all
square integrable. 
Therefore, in this case,  the single-particle level density  is 
 given by  Eq.~(\ref{discreteleveldens}). By introducing 
 the Green's operator, $\hat G_{\infty}(z)=(z-\hat H_{\infty})^{-1}$,
 $g_{\rm d}(\epsilon)$ can be written as
\begin{equation}\label{dgreen}
g_{\rm d}(\epsilon)=-{1\over \pi}{\rm Im}\left\{ {\rm Tr}\left[\hat
G_{\infty}(\epsilon)\right]\right\}.
\end{equation}

As discussed in  Ref.~\cite{[Kru98]},  for a Hamiltonian $\hat H$ containing
a finite potential
the full level density  (\ref{e3}) becomes
\begin{equation}\label{discretgreen}
g(\epsilon)=-{1\over \pi}{\rm Im}\left\{ {\rm Tr}\left[\hat G^+(\epsilon)-
\hat G_{0}^+(\epsilon)\right]\right\},
\end{equation}
where  $\hat G^+(z)=(z-\hat H+i0)^{-1}$ and  $\hat G^+_0(z)=(z-\hat H_0+i0)^{-1}$ is the
free outgoing Green's operator associated with  $\hat H_0=\hat T$.
The interpretation of Eq.~(\ref{discretgreen}) is straightforward:
the  second term contains the contribution to the
single-particle level density originating from the gas of free particles.

Let us now  introduce an approximation
to the  exact expression  (\ref{discretgreen}). To this end,
we  diagonalize  $\hat H$ and $\hat H_0$  in an orthonormal
basis formed from the $M$ square-integrable basis functions.
The resulting  approximate eigenenergies of  $\hat H$ 
and $\hat H_0$ are denoted by
$e_i$ and    $e^0_i$, respectively  ($i$=1,...,$M$). 
This procedure amounts to a projection of both Hamiltonians into 
the  $M$-dimensional Hilbert space of square-integrable basis functions.
 The  level density 
(\ref{e3}) can  then be approximated by the difference of the
discrete level densities of the two projected Hamiltonians:
\begin{equation}\label {gM}
g_M(\epsilon)= \sum_{i=1}^M 2 \delta(\epsilon -e_i) -  \sum_{i=1}^M 2 \delta(\epsilon -e^0_i).
\end{equation} 
By increasing the dimension  $M$,  the bound eigenvalues  of $\hat H$  converge to the exact 
single-particle energies while the  positive-energy  eigenvalues will tend to approach 
zero energy. The  eigenvalues $e_i^0$, which   are obviously different from  
the positive energies $e_i$  for any finite $M$, can, in fact, compensate for the spurious
 increase of the level density around the zero energy if the smoothing procedure (\ref{ggsmooth})
 is applied:
\begin{equation}\label{gMsmooth}
\tilde{g}_M(\epsilon) = {1\over\gamma}
\int_{-\infty}^{+\infty}d\epsilon'~ g_M(\epsilon')
f\left(\frac{\epsilon'-\epsilon}{\gamma}\right).
\end{equation}
It has been shown in Ref.~\cite{[Kru98]} that the  exact
smoothed level density $g(\epsilon)$ can be reproduced by $\tilde{g}_M(\epsilon)$ 
 in the limit of large   $M$. 

In this work we employ the (stretched)  harmonic oscillator basis. As discussed below,
thirty oscillator shells are sufficient to guarantee the convergence of results.
In the standard (old) method, one takes much fewer states (12-14
 oscillator shells)   and the second term in Eq.(\ref{gM}) is not
 subtracted. Consequently, the results are not stable 
as $M$ is varied.

\subsection{Generalized Shell-Correction Method}\label{newtreatment}

In the standard Strutinsky smoothing method, when applied to finite potentials,
it is difficult to meet
the plateau condition (\ref{plateau})   \cite{[Naz94]}. The more detailed study of  Ref.~\cite{[Ver98]}
 demonstrated that  the plateau condition can 
seldom be satisfied even if the particle  continuum
is properly accounted for. Fortunately,  it is possible to replace
the standard plateau condition with a new requirement  which yields
unambigous shell-correction values \cite{[Ver98]}. By comparing the smoothed
Strutinsky level densities with those obtained  in  the semiclassical
 Wigner-Kirkwood method, it was found
that they are in good agreement,
apart from the low and the high ends of the spectra. In the intermediate energy region,
the average level density 
shows linear dependence on  $\epsilon$.
(The linearity of the semiclassical level density for heavy nuclei
was noticed already in Ref.~\cite{[Shl92]}.) 
Guided by this observation,  the shell-correction method was generalized by
replacing the plateau condition with  the requirement
that in an energy interval $[\epsilon _l,\epsilon _u]$ which is wider than the average
 distance between neighboring major shells,  
\begin{equation}\label{sav}
\epsilon _u - \epsilon _l = 1.5~\hbar\omega_0,
\end{equation}
the deviation of $g(\epsilon,\gamma,p)$ from linearity should be minimal \cite{[Ver98]}.
In practice, one has to minimize the deviation
\begin{equation}\label{khi2}
\chi ^2(\gamma,p)=\int_{\epsilon_l}^{\epsilon_u}\left[g(\epsilon,\gamma,p)-a-b\epsilon\right]^2
d\epsilon,
\end{equation}
where the parameters $a$ and $b$ are uniquely determined for each value of  $\gamma$
and $p$ by  the method of least squares.

Figure~\ref{Fig1} displays $\chi^2$  as a function
of $\gamma$ and $p$ for the neutrons in  the spherical superheavy nucleus $Z$=114,  $N$=184.
It is seen that  $\chi ^2$ has  two minima  for each $p$ value, and there is a  clear
correlation between $p$ and $\gamma$ (minima are shifted to larger  values of $\gamma$
with increasing $p$).  The position of the first minimum is between
1.12 and 1.48  $\hbar\omega_0$  (i.e.,  between 6.6 MeV and 8.6 MeV).
The second minimum appears  at larger $\gamma$ values, namely between 7.7 MeV and 12
MeV. As demonstrated in Ref.~\cite{[Ver98]},  the average level densities
corresponding to the first minimum of $\chi^2$ practically do not depend on $p$
in the negative energy region. This is  also valid for 
the second minimum in $\chi^2$. 
However, a  difference can be seen if one compares 
average  level densities 
calculated at  different minima.  The inset of Fig.~\ref{Fig1}
shows $\tilde{g}(\epsilon)$ for both minima at $p$=10.
The level density corresponding to the lower-$\gamma$ minimum
preserves its linearity  for
a wider range of energies,  and
the linearity  is best fulfilled in the energy region
which is midway   at the bottom and the top of the potential.
Therefore,  in our calculations [Eq.~(\ref{khi2})]   we fixed the energy
interval so that it is centered
around  the half of the energy of the lowest single-particle level.
Far from this central region  ${g}(\epsilon)$ varies rapidly
and this  forces the smoothed level density to oscillate.  These
oscillations can be viewed as  {\em end effects},  and they are 
largely independent of the  shell 
structure. For example,  for the
harmonic oscillator potential whose  spectrum has no natural upper bound, 
these  oscillations occur around the bottom of the potential well. For 
the Woods-Saxon potential, additional oscillations occur around 
the threshold energy.

   It is well known that the realistic value of the smoothing parameter $\gamma$
has to lie in a certain energy interval \cite{[Jen74]}.
The value of $\gamma$  should be large enough
to wipe out shell effects in the energy  range of
a typical distance between shells,
but it  should not be much larger  to avoid bringing the
threshold oscillations down to 
lower energies. To prevent this, in our calculations we always use  the
$\gamma$ values corresponding  to the 
first minimum of  $\chi ^2$.
For the case  shown in Fig.~\ref{Fig1},   shell correction
 changes by 0.3 MeV if one  uses the value of  $\gamma$ at the second
minimum instead. While in this case the change in shell correction 
 is well within the uncertainty of the model,
the difference is  more pronounced
for lighter
nuclei.
For the very light nuclei,  the 
   end effects dominate the energy dependence of 
 $\tilde g(\epsilon)$ in
the whole energy range.
Consequently,  the Strutinsky smoothing method
 cannot be meaningfully applied to these systems.

\section{Results}\label{results}

\subsection{Model Parameters}

In the calculations presented in this work,  we have used the average,
axially deformed   Woods-Saxon
(WS) potential of Ref.~\cite{[Cwi87]}, which contains the  central part, 
the  spin-orbit term,
and the
Coulomb potential for protons.
The potential depends  on a set of
deformation parameters, $\beta_\lambda$, defining the nuclear surface:
\begin{equation}\label{Ro}
R(\theta; \bbox{\beta}) = C(\bbox{\beta}) r_\circ A^{1/3} \left[ 1 +
\sum_{\lambda} \beta_{\lambda} Y_{\lambda 0} (\theta)
\right],
\end{equation}
where the coefficient $C$ assures that the total volume enclosed by
the surface  (\ref{Ro}) is conserved.
The Coulomb potential has been  assumed to be that of the charge
$(Z-1)e$ uniformly distributed within the deformed nuclear surface.
 We employed the set of WS
parameters introduced in Ref.~\cite{[Dud81]}. For details
 pertaining to the
WS  model,  see Refs.~\cite{[Naz94],[Cwi87]}.

The deformed WS Hamiltonian was diagonalized in the deformed harmonic oscillator
basis using the computer code of  Ref.~\cite{[Cwi87]}.
For the diagonalization we took
all the (stretched) oscillator states having the principal
 quantum number less or equal than  $N_{\rm max}$.
(In short, we took  $N_{\rm max}$ deformed shells.)
The diagonalization of  $\hat H_0$ was carried out in precisely the same basis.

When adopting the new scheme,
the important question  is how many harmonic oscillator shells
are needed  in order to reproduce the  exact  results of Ref.~\cite{[Ver98]}.
 Naturally,  the value of $N_{\rm max}$  depends on the size and the shape of the potential
to be diagonalized, and also on the oscillator frequency
$\hbar \omega=\eta \hbar\omega_0$ (the default value of
$\eta$ is 1.2).

The convergence of $\delta E_{\rm shell}$ for neutrons
as a function of $N_{\rm max}$ is illustrated in
Fig.~\ref{Fig2} for  $^{132}$Sn   and
 $^{154}$Sn (which is weakly bound).
  One can see that for both nuclei the shell correction obtained in the
  new procedure  quickly converges to 
  the exact value,  and at  $N_{\rm max}$=30 
the agreement is very satisfactory. Therefore, in the following, we shall    use 
30 oscillator  shells when applying the new method. 

As expected, the results of calculations using
the standard method do  not stabilize with $N_{\rm max}$.  
However, for the recommended values of $N_{\rm max}$=12 and 14,
the shell correction  produced with the old method differs from the exact value 
by less than 1 MeV. 
 However this apparent agreement seems to be accidental.  We display
in Fig.~\ref{Fig2}c and \ref{Fig2}d, respectively,   the single-particle
energy  and the smoothed single-particle energy  for $^{154}$Sn  as a function of  $N_{\rm max}$. One can 
notice that  at $N_{\rm max}$=12 the total single-particle energy  differs from the  exact value
by about 8 MeV. Since the corresponding smoothed single-particle energy 
is also shifted by about 8.5 MeV,  the resulting shell correction 
differs only by 0.8 MeV from the exact value. 
Consequently, an acceptable agreement for 
$\delta E_{\rm shell}$ comes as a result of cancellation between
  two large numbers, each  subject to large errors.
As   one approaches   the neutron drip line,  the accurate calculation of 
single-particle energies requires   a rather high number of shells and/or a
basis optimization with respect to the  parameter $\eta$ which determines the oscillator length.
This is illustrated in
 Fig.~\ref{Fig3} which shows the convergence of the  total neutron
 single-particle energy for $^{120}$Zr at 
a  large quadrupole deformation $\beta_2$=0.6.  Clearly, for a weakly bound and deformed
nucleus one needs at least $N_{\rm max}$=30 oscillator shells to reach
the  convergence with the
standard value of $\eta$.
Of course, by increasing the oscillator length, i.e., by 
choosing a smaller value of $\eta$, one can improve the convergence
significantly for a system with a spatially extended density. 
In this example,  one can arrive at a  reasonably
accurate value  of $E_{\rm s.p.}$
by using   $N_{\rm max}$=20 and $\eta$=0.8. Another practical way of improving
the convergence of 
single-particle energies is to use the modified oscillator
 basis  obtained  by means of the local scaling  transformation
\cite{[Sto98]}.

Figure~\ref{Fig4}  shows  the neutron smoothed level density for
$^{132}$Sn, a relatively well bound nucleus, calculated with different methods.
The new method with   $N_{\rm max}$=30 describes very well $\tilde g(\epsilon)$  in the
whole region of negative energies. This proves  that the Green's function approach
can be used with confidence, even for weakly bound systems. On the other hand,
the average level density obtained with 
the standard  
method  never stops   increasing,
  and its deviation from the exact result shows up already at $\epsilon$=--18 MeV. 
 The result  displayed in Fig.~\ref{Fig4} demonstrates that even for well-bound nuclei,  
  the shell corrections calculated
using the old method are prone to significant errors. 

\subsection{Deformation Effects}

In order to investigate the deformation dependence of shell corrections, 
we performed calculations for 
$^{100,110,120}$Zr as a function of $\beta_2$
(other deformation parameters were assumed to be zero).
In the Green's function variant, the generalized plateau condition was used;
the resulting values of $\gamma$ were also employed  in the 
standard Strutinsky calculations.
The results are shown in Fig.~\ref{Fig5}. As  expected, the 
most pronounced  difference
between the results of the two methods is  for the weakly 
bound nucleus,  
$^{120}$Zr. This difference does depend on $\beta_2$; part of it
can be attributed to the deformation dependence of the smoothing width.
(It should be noted that in the deformed calculations of Ref.~\cite{[Naz94]}
$\gamma$ was assumed to be constant.)

Since the effect of the particle continuum on $\delta E_{\rm shell}$
 should be less pronounced for  systems that are bound better, one would expect
the  two methods  to yield similar results for lighter
Zr isotopes. However, as seen in Fig.~\ref{Fig4}, the difference between 
both methods is negligible only at very low energies, $\epsilon$$<$18\,MeV.
For higher values of $\epsilon$ (or Fermi level  $\tilde\lambda$),  the
difference between the smoothed level densities 
is not negligible and it is not even a monotonous  function
of $\tilde\lambda$.   As a result,  one can notice in Fig.~\ref{Fig5} that
for $^{110}$Zr the results of both methods are very close while in a better bound
nucleus of $^{100}$Zr they differ more.
It is interesting to note that  for $^{100}$Zr and $^{110}$Zr the deformation 
dependence of   $\delta E_{\rm shell}$ is very similar in both variants
of calculations. This is in accord with the observation made in Ref.~\cite{[Naz94]}
that the  difference between the shell corrections 
obtained in the standard method and the
semiclassical approach
depends rather weakly on deformation.

As discussed earlier,  two main sources of the error of the standard smoothing
procedure are (i) the error in determining  the total single-particle energy, and
(ii) the uncertainty of smoothing that influences the value of  the
smooth single-particle energy. Since (i) and (ii)  are not independent, a large cancellation takes
place which might reduce the total error.
However,  it is {\em a priori} difficult to predict how large
this difference is
and how strongly it  would  affect the predicted position of the drip line.
In order to shed some light on this question,  we 
carried out a comparison between  shell corrections calculated
with the two methods for the spherical Sn and $Z$=114 isotopes, and
for the deformed Zr and Er isotopes. (In the latter case, we fixed deformation
at $\beta_2$=0.2.) These nuclei
 represented medium-mass
and heavy nuclei where the generalized shell-correction method can be applied.
The results are presented in Fig.~\ref{Fig6}.
Except for well-bound Zr and $Z$=114 isotopes,  the difference
between the shell corrections calculated using the old and the new methods
are on the order of MeV, and, except for the Sn isotopes, it is
rather large when approaching
  $\tilde\lambda=0$ (drip line). 
 Although our
 calculations do not aim at   determining the  actual position
of the drip line, they give  a reasonably good estimate for
the uncertainty of the old procedure. If one  identifies the drip line 
with the neutron number where  $\lambda$ becomes positive
(this assumption is usually violated in 
actual  calculations
because of the lack of self-consistency
between the microscopic and macroscopic parts of the energy formula \cite{[Naz94]}),
our limited calculations suggest that the drip-line predictions by the standard
method  are prone to severe uncontrolled uncertainties.

\subsection{Modification  of the Green's Function Method to the Proton Case}

 In the presence of the long-range Coulomb potential, the free 
Hamiltonian appearing in Eq.~(\ref{discretgreen}) has to be modified. Indeed, 
for the protons, the 
asymptotic
behavior of the scattering states is that of the Coulomb functions, not  plane waves.
Therefore, in this case, for the free Hamiltonian we take
\begin{equation}\label{freecoul}
 \hat{H}_0=\hat T+V_{\rm Coul},
\end{equation}
with  $V_{\rm Coul}$ being  the Coulomb potential.
The role of the Coulomb term is to effectively push the 
continuum  up in energy to the top of the Coulomb barrier. The results are insensitive to the radius of the
Coulomb potential in the free Hamiltonian. As a matter of fact, even a point Coulomb potential 
can be used in Eq.~(\ref{freecoul})  \cite{[Ara99]}.

Figure~\ref{Fig7} shows   the smoothed proton level density 
in the proton-rich nucleus $^{180}$Pb.  One can see that 
already with a rather  low value of  $N_{\rm max}$=19
the new method 
 reproduces the exact
smoothed level density in the whole range of negative energies,
and $N_{\rm max}$=30 gives an excellent agreement with  the exact result.
The reason that relatively low values of $N_{\rm max}$ are  sufficient
 in the proton case
is that even slightly unbound states (narrow proton resonances) are well localized due to the 
confining effect of 
the Coulomb barrier.

\section{Conclusions}\label{conclusions}

In 
this work, we employed the Green's function oscillator
expansion  method to calculations of shell corrections.
For spherical nuclei, the new method has proved to be a
fast and very accurate  approximation to the exact procedure.
It also allows for a straightforward  generalization to
deformed shapes. In essence, the method is based on two
simultaneous diagonalizations in a large oscillator basis.
The first diagonalization involves the actual one-body Hamiltonian
while the other one is carried out for the free Hamiltonian representing
the particle gas 
whose contribution to the level density should be subtracted.
For the neutrons, the free Hamiltonian is given by the kinetic energy
operator, while for the protons
it also includes the Coulomb potential.
In practice, the space of 30 (stretched) oscillator shells is sufficient
to guarantee the stability of  results.
This relatively large (but still tractable)
space is necessary not only for the proper 
treatment of the free gas
 but also for the  accurate calculations the total single-particle energy.

As demonstrated
in our study, the use of 
the standard smoothing procedure
can lead to serious deviations when extrapolating off beta stability.
In particular, the    particle 
drip lines predicted in the traditional approach  can be very uncertain.
(The systematic error 
in $\delta E_{\rm shell}$, 
due to the particle continuum,
can be as large as several MeV at the neutron drip line.)
According to our calculations,  the error on $\delta E_{\rm shell}$
depends weakly on deformation in most cases. It is only
for the weakly bound nuclei that the difference between the old and new methods
exhibits a sizable deformation dependence.

There is no simple ``fix" that would cure the 
deficiencies of the standard Strutinsky procedure when applied to
finite-depth potentials. One does
need  the large basis in order to guarantee the stability
of $E_{\rm s.p.}$. On the other hand, at these large values of $N_{\rm max}$, the smooth
single-particle energy becomes unreliable due to the unphysical increase of the 
quasi-bound levels  around the threshold.
We  believe that the new Green's function 
method, together with the generalized plateau condition,
is a very useful  tool that should be employed  in 
future global calculations
of nuclear masses in the framework of the one-body 
(macroscopic-microscopic)
description and  in level-density calculations for spherical
and deformed nuclei. Of course, the new procedure does not
remove the generic problem of the lack 
 of the
self-consistency condition between the microscopic and macroscopic
Fermi energies \cite{[Naz94]}. Recently, the Green's function
method,  based on self-consistent 
potentials obtained in Hartree-Fock and relativistic mean-field
calculations,  was used  to extract
shell corrections in the spherical superheavy nuclei \cite{[Kru00]}. 
Although these calculations were not done using the oscillator
basis expansion method but directly in the coordinate space,
their main principle  is the same as that discussed in  this paper.

\acknowledgments

This research was supported in part by
 the Hungarian National Research Fund
(OTKA T026244 and T029003) and 
 the U.S. Department of Energy
under Contract Nos. DE-FG02-96ER40963 (University of Tennessee),
DE-FG05-87ER40361 (Joint Institute for Heavy Ion Research), and 
DE-AC05-96OR22464 with Lockheed Martin Energy Research Corp. (Oak
Ridge National Laboratory).


\newpage
\newlength{\myplotsize}
\setlength{\myplotsize}{16cm}

\begin{figure}
\vbox{
\begin{center}
\leavevmode
\epsfxsize=\myplotsize
\epsfbox{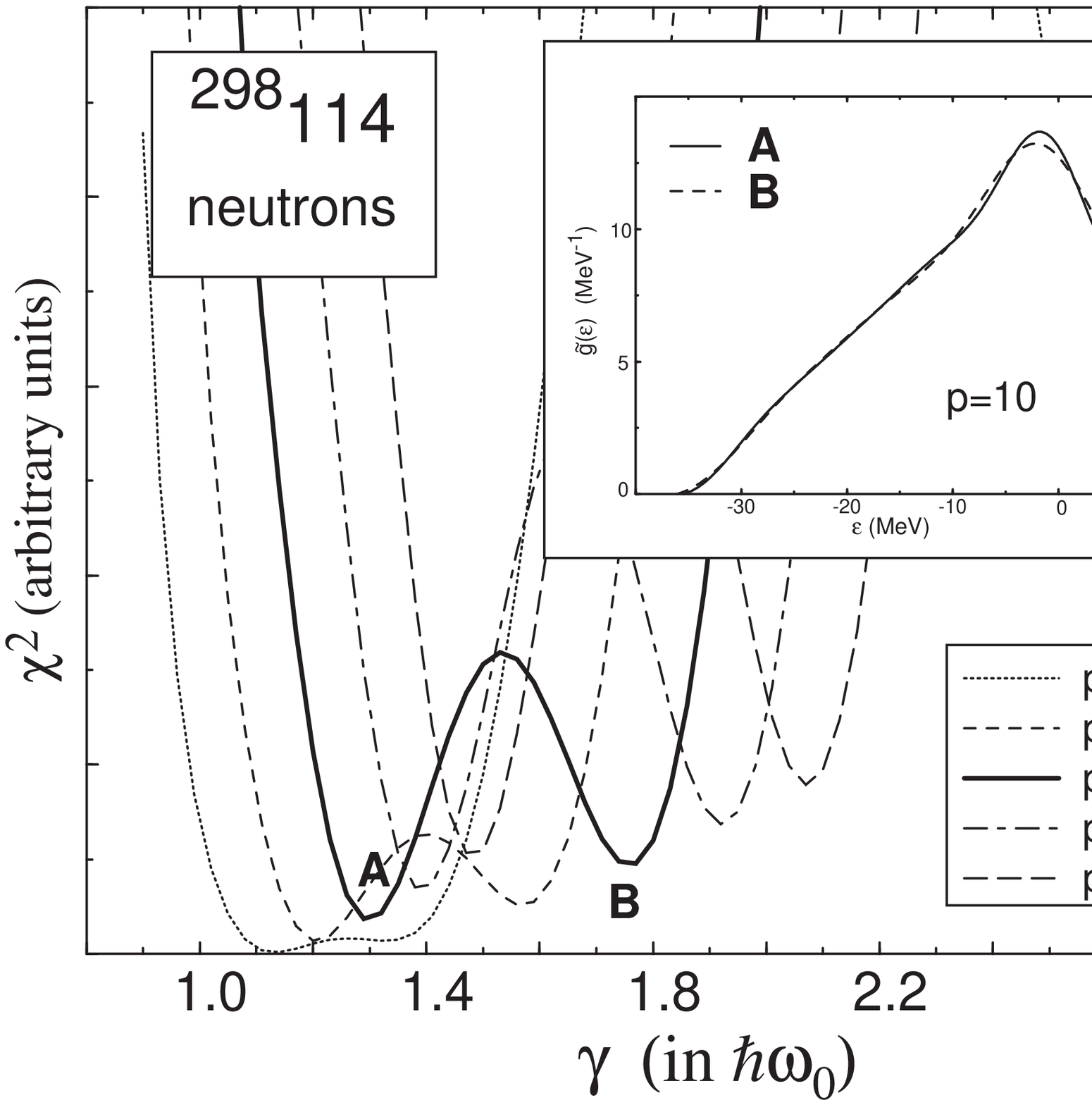}
\end{center}
\caption{Dependence of $\chi^2$ of Eq.~(\protect\ref{khi2}) on the smoothing range $\gamma$
for the spherical superheavy nucleus $Z$=114, $N$=184. The calculations were performed
for several values of the curvature order $p$.
The inset shows the average level
densities corresponding to the two minima A and B of  $\chi^2$ with $p$=10.
}
\label{Fig1}
}
\end{figure}

\begin{figure}
\vbox{
\begin{center}
\leavevmode
\epsfxsize=\myplotsize
\epsfbox{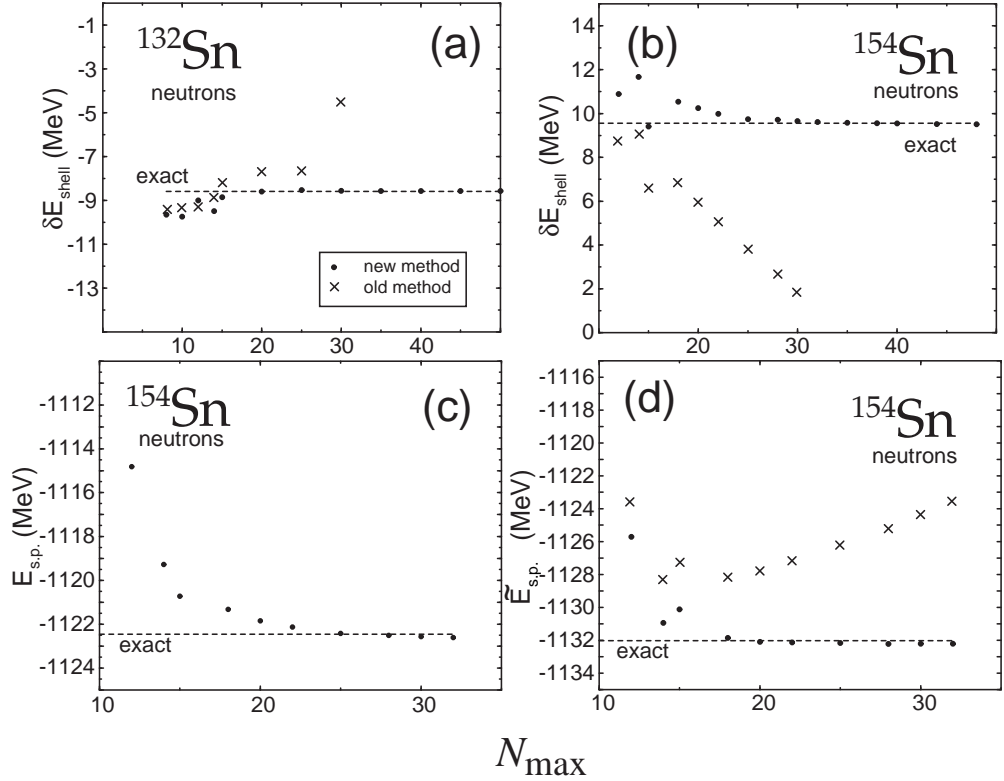}
\end{center}
\caption{Dependence of the  neutron shell correction on the size of the harmonic
oscillator basis  for (a) $^{132}$Sn and for (b) $^{154}$Sn. All 
of the single-particle
states with  principal oscillator quantum number  less or equal than 
$N_{\rm max}$ were considered  in the diagonalization.
The $N_{\rm max}$-dependence of  $E_{\rm s.p.}$ and  $\tilde E_{\rm s.p.}$
for  $^{154}$Sn is shown in portions (c) and (d), respectively.
}
\label{Fig2}
}
\end{figure}

\begin{figure}
\vbox{
\begin{center}
\leavevmode
\epsfxsize=\myplotsize
\epsfbox{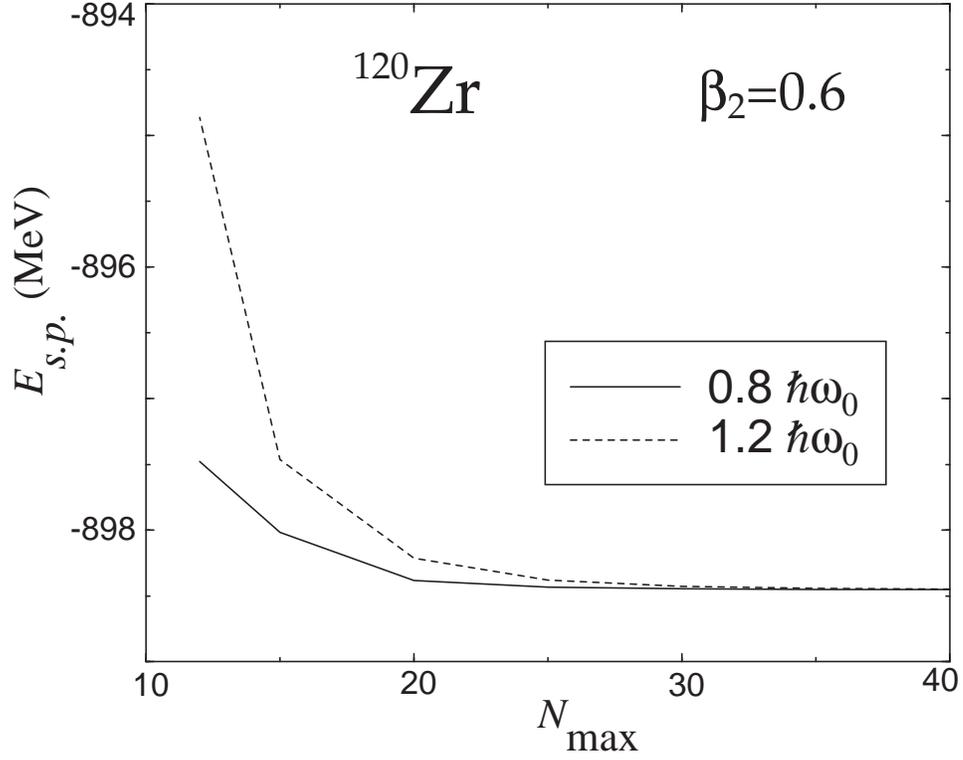}
\end{center}
\caption{Dependence of the total single-particle energy for 
$^{120}$Zr at $\beta_2$=0.6 on the size of the
stretched harmonic oscillator  basis $N_{\rm max}$ for the
two values of the oscillator frequency parameter
($\hbar\omega/\hbar\omega_0$=0.8 and 1.2, where  $\hbar\omega_0=41/A^{1/3}$ MeV).
}
\label{Fig3}
}
\end{figure}

\begin{figure}
\vbox{
\begin{center}
\leavevmode
\epsfxsize=\myplotsize
\epsfbox{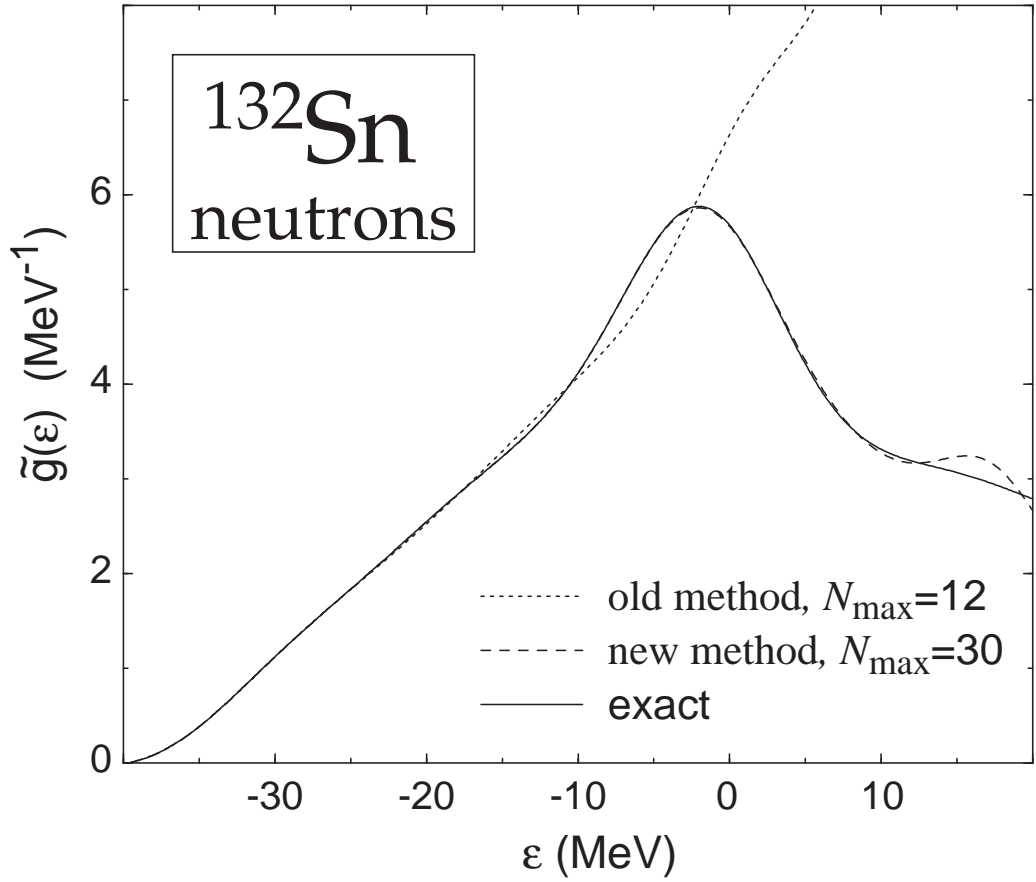}
\end{center}
\caption{Neutron average  level density for the spherical nucleus $^{132}$Sn
calculated  using the
Green's function  method with $N_{\rm max}$=30
(dashed line),  the standard smoothing  method  with $N_{\rm max}$=12
(dotted line), and  the Gamow-state technique 
(exact result, solid line).
}
\label{Fig4}
}
\end{figure}

\begin{figure}
\vbox{
\begin{center}
\leavevmode
\epsfxsize=\myplotsize
\epsfbox{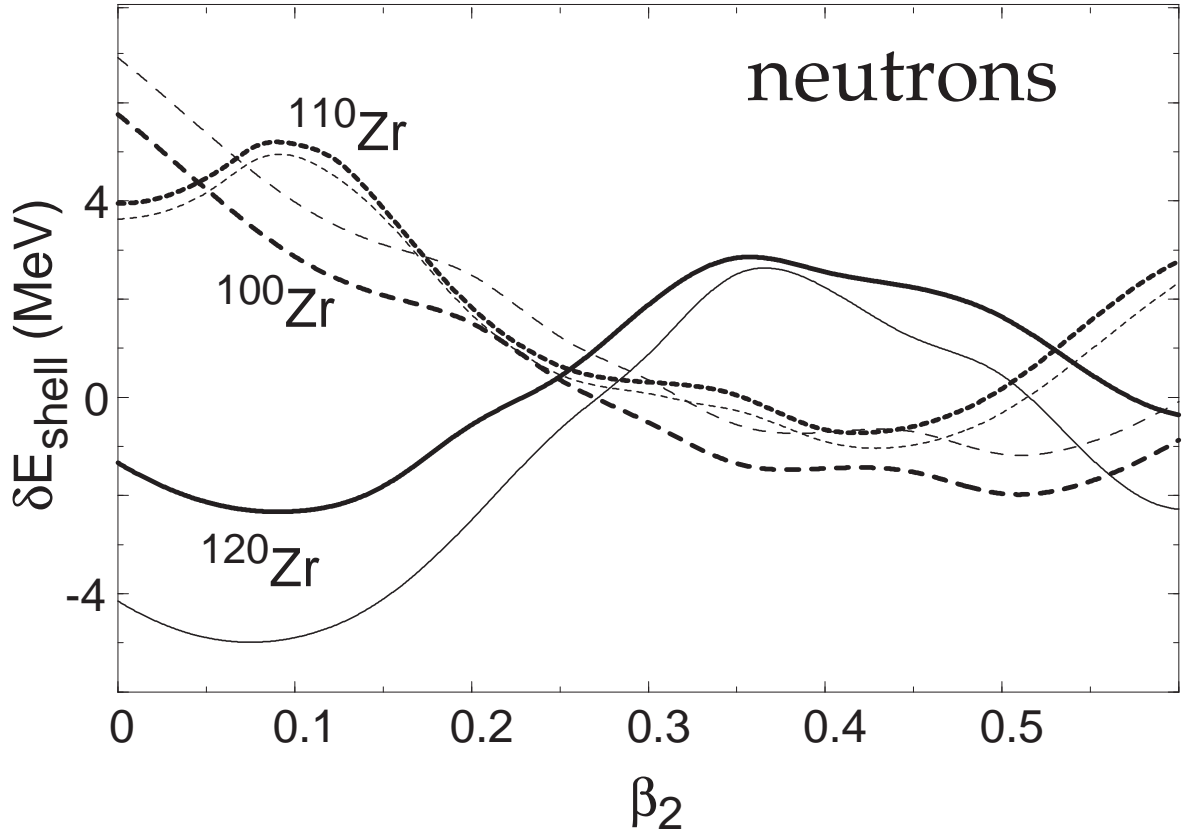}
\end{center}
\caption{Neutron shell corrections  for  $^{100}$Zr (short dashed lines),
$^{110}$Zr (dashed lines),  and $^{120}$Zr (solid lines). The results of the
Green's function  method with $N_{\rm max}$=30
are shown by thick lines while  those of the  standard smoothing  method  with $N_{\rm max}$=12
are indicated by thin lines.
}
\label{Fig5}
}
\end{figure}

\begin{figure}
\vbox{
\begin{center}
\leavevmode
\epsfxsize=\myplotsize
\epsfbox{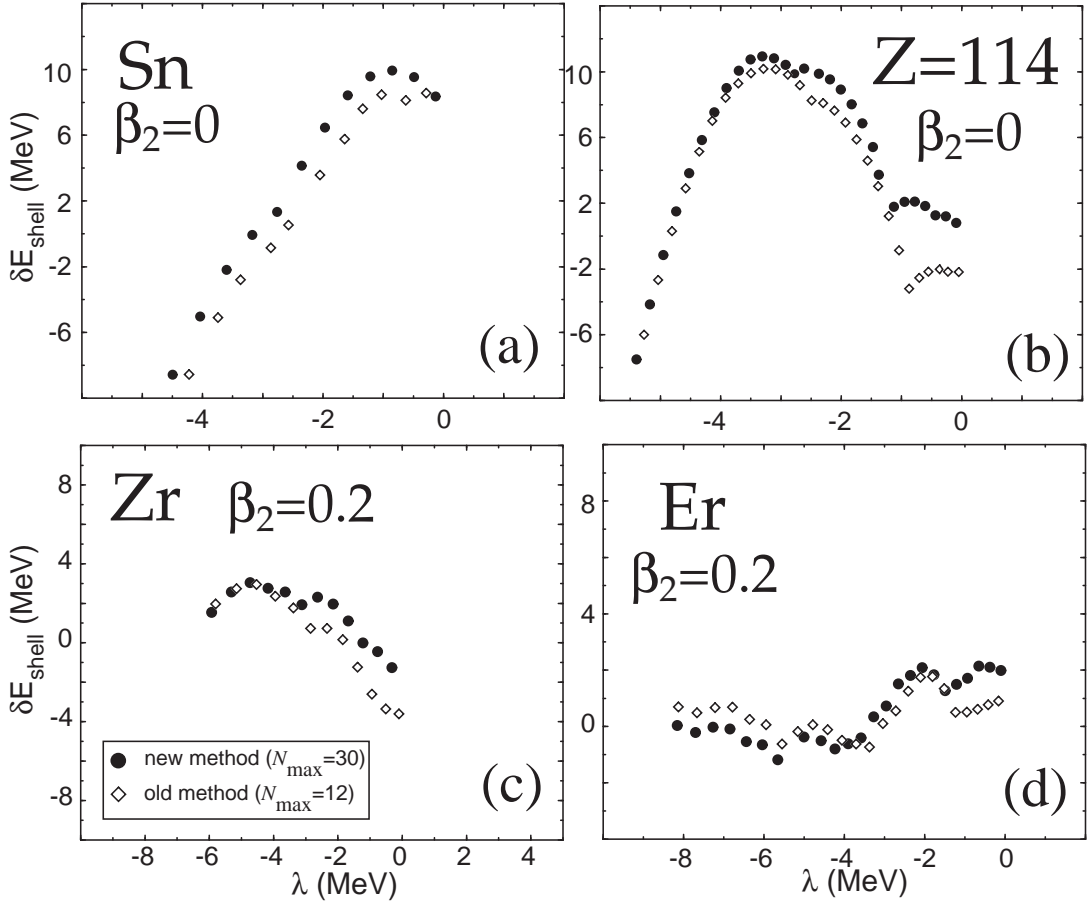}
\end{center}
\caption{Neutron shell corrections  for  
spherical Sn and $Z$=114 nuclei and deformed ($\beta_2$=0.2) 
Zr and Er isotopes as a function of the neutron Fermi level $\lambda$.
The Green's function  and standard calculations are shown by filled
and open symbols, respectively.
}
\label{Fig6}
}
\end{figure}

\begin{figure}
\vbox{
\begin{center}
\leavevmode
\epsfxsize=\myplotsize
\epsfbox{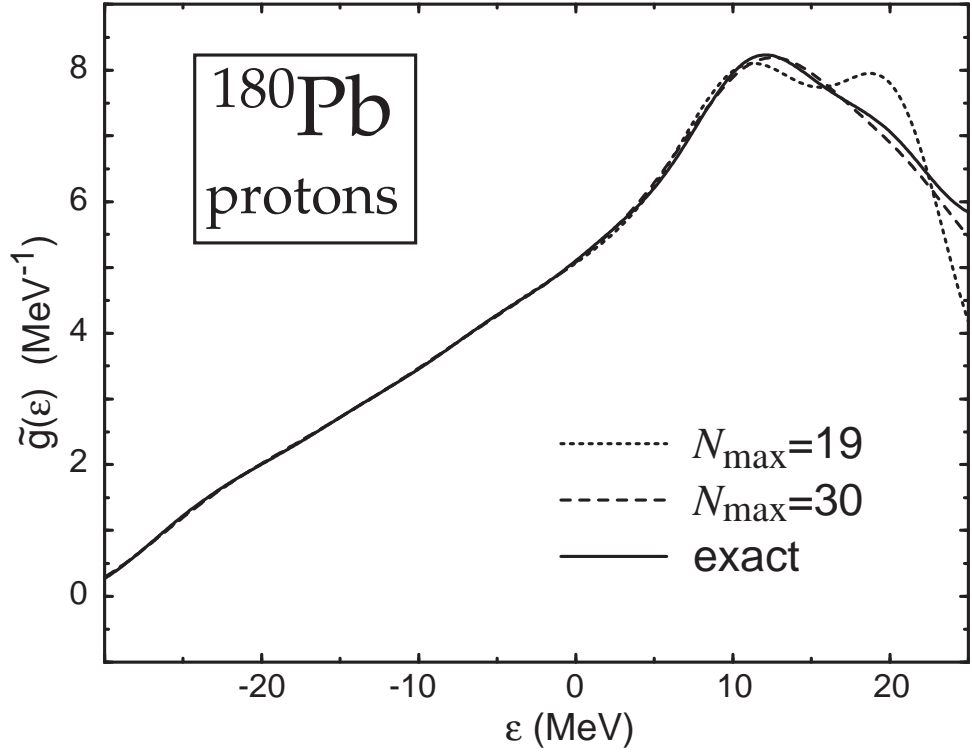}
\end{center}
\caption{Proton  average  level density for the spherical nucleus $^{180}$Pb
calculated  using the
Green's function  method with $N_{\rm max}$=30
(dashed line) and  $N_{\rm max}$=19 (dotted line),
and  the Gamow-state technique 
(exact result,  solid line).
}
\label{Fig7}
}
\end{figure}

\end{document}